\documentclass[reprint,showpacs,prl,superscriptaddress]{revtex4-1}
\usepackage{epsfig}
\usepackage{amsmath}

\usepackage{graphicx}  
\usepackage{dcolumn}   
\usepackage{bm}        
\usepackage{amssymb}   
\usepackage{amsmath}
\begin{document}


\title{The Classical-to-Quantum Transition with Broadband Four-Wave Mixing}
\author{Rafi Z. Vered, Yaakov Shaked, Yelena Ben-Or, Michael Rosenbluh and Avi Pe'er}
\affiliation{Physics Department and BINA Center for Nano-technology, Bar-Ilan University, Ramat-Gan 52900, Israel}
\email{avi.peer@biu.ac.il}

\begin{abstract}
A key question of quantum optics is how nonclassical bi-photon correlations at low power evolve into classical coherence at high-power. Direct observation of the crossover from quantum to classical behavior is desirable, but difficult due to the lack of adequate experimental techniques that cover the ultra-wide dynamic range in photon flux from the single photon regime to the classical level. We investigate bi-photon correlations within the spectrum of light generated by broadband four-wave mixing (FWM) over a \emph{large dynamic range of $\sim80dB$ in photon flux} across the classical-to-quantum transition using a two-photon interference effect that distinguishes between classical and quantum behavior. We explore the quantum-classical nature of the light by observing the interference contrast dependence on internal loss and demonstrate quantum collapse and revival of the interference when the FWM gain in the fiber becomes imaginary.
\end{abstract}
\maketitle

In FWM, two pump frequencies $\omega_1, \omega_2$ are converted into another pair of frequencies $\omega_s, \omega_i$, such that $\omega_1\!+\!\omega_2\!=\!\omega_s\!+\!\omega_i$ (energy conservation). Quantum mechanically, FWM represents the simultaneous annihilation of a pair of pump photons and the creation of another pair of signal-idler photons (bi-photon). The prevalence of FWM in practically any medium, and specifically in optical fibers, renders it most important for applications, such as  parametric amplifiers, oscillators \cite{Hasegawa1980tunable,Nakazawa1988Modulational,Hansryd2002application} and frequency-comb sources \cite{okawachi2011octave, ferdous2011spectral,johnson2012chip,ferdous2012probing}. Spontaneous FWM is an important source of single bi-photons for quantum information and quantum communications schemes, especially for in-fiber applications \cite{fiorentino2002all,li2005optical,soeller2011high}.

Classically, FWM is a phase-dependent gain process that amplifies one quadrature of the combined signal-idler input field and de-amplifies the other quadrature \cite{Stolen1983Parametric}. As a pure gain mechanism, classical FWM does not incorporate spontaneous bi-photon emission, and an input seed is necessary in the classical model to initiate the nonlinear conversion. To introduce spontaneous emission, one must add a fictitious white noise seed, which acts as a classical representation of vacuum fluctuations - a semi-classical treatment that is adequate when the parametric gain is high and the nonlinear conversion overwhelms the fictitious seed \cite{abram1986direct, dayan2004two, Vered2012Two}. However, for low parametric gain, the semi-classical model fails and a quantum treatment is necessary. In the limit of very low gain, a perturbative treatment is valid, where multiple-photon states are neglected, describing the regime of spontaneously generated single bi-photons, as commonly used in quantum optics \cite{fiorentino2002all,li2005optical,soeller2011high}.

Our work focuses on the intermediate gain regime, which is more challenging, as it requires complete quantum treatment of the field, incorporating both spontaneous and stimulated emission. This regime, at the crossover from the quantum to the classical, is characterized by the emergence of classical coherence out of the quantum bi-photon correlation. It is especially relevant for generation of frequency combs by FWM in micro-cavities \cite{wang2012observation}, where the broadband phase-coherence properties are critical, but not yet well understood \cite{herr2011universal, weiner2011FWMcomb}. Specifically, it is important to observe how, and under what conditions broadband, classical phase-locking emerges from single photon quantum correlation \cite{foster2011silicon,herr2011universal,weiner2011FWMcomb,wang2012observation,pasquazi2012stable,delhaye2014self}. Intuitively, the crossover between quantum and classical behavior occurs when the average photon flux exceeds one photon-pair per correlation-time of the bi-photons \cite{dayan2005nonlinear, dayan2007theory} ($\sim 25 $fs in our FWM experiment \cite{Vered2012Two}). When the bi-photon flux is lower than this limit, the FWM process is primarily spontaneous and its properties are only predicted by quantum mechanics, whereas for higher flux, stimulated FWM becomes pronounced and the quantum predictions converge to the classical analysis.

Experimentally, this intermediate regime is unexplored due to the lack of adequate detection methods. The wide dynamic range of light intensity from the single photon pair level to the multi-mW range cannot be covered, neither by single photon counting and coincidence circuits, which saturate at $10^6\!-\!10^7$ photons/s, nor by photo-detectors, which are insensitive to low photon flux \cite{thomas2012practical}. Furthermore, in order to measure the optical phase, a local-oscillator (LO) is necessary as a phase reference for homodyne detection. Such an LO is difficult to obtain for FWM, especially for broadly separated, correlated frequency pairs, which require two \emph{phase-correlated} LOs. So far, such LO-pairs could be obtained only by seeding the FWM process at a specific frequency \cite{boyer2008entangled} or by delicate referencing to optical cavities \cite{Villar2006Direct}.

\begin{figure}
\includegraphics[width=8.6cm]{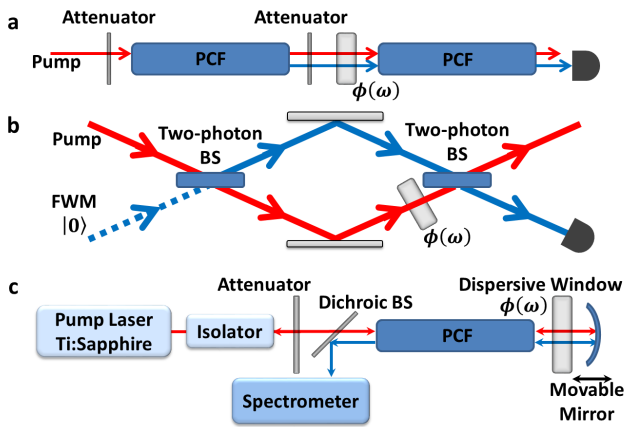}
\caption{\label{fig:setup} Experimental concept and setup. (a) The concept: two FWM media with broadband phase-matching are placed in series and pumped by a narrowband laser. In between, a dispersive element introduces a spectrally dependent phase between the pump and the FWM bi-photons, causing nonlinear interference fringes to appear on the intensity of the FWM spectrum generated in the second fiber. Quantum mechanically, the experiment is equivalent to a two-photon Mach-Zehnder interferometer, as shown in (b), where the PCFs represent bi-photon beam-splitters that couple between the pump and the signal-idler pairs. (c) The experimental configuration: 6 ps pump pulses near 785 nm (Tsunami by Spectra Physics with $\sim80 $MHz repetition) enter a 12 cm long photonic crystal fiber (PCF, zero dispersion at ~784 nm), generating a broadband spectrum of signal-idler pairs. After the first pass through the fiber, the pump and the signal-idler pairs are reflected back for a second pass through the fiber and the resulting FWM spectrum is separated from the pump (by a dichroic filter) and measured with a high resolution spectrometer coupled to a cooled, intensified CCD camera. In between the passes, a dispersive window creates a spectrally dependent phase. Attenuation of both the pump and the FWM sidebands between the two passes is controlled by a variable neutral-density filter (not shown).}
\end{figure}

\begin{figure} 
\includegraphics[width = 8.6cm]{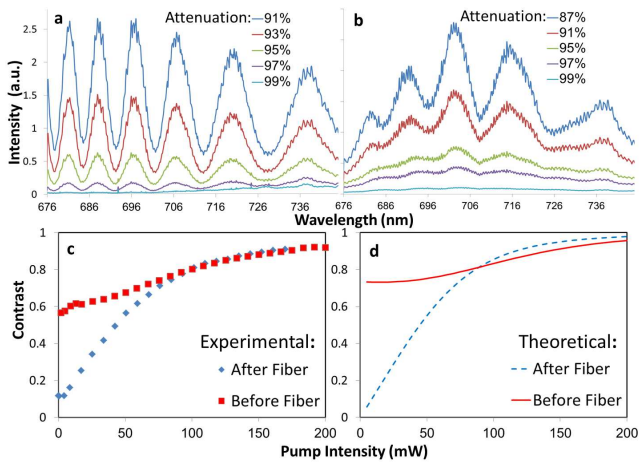}
\caption{\label{fig:contrast784nm}  (a) Measured spectral interference when attenuation is applied before the fiber. (b) The same interference when attenuation is applied after the first pass through the fiber. Both graphs present the same five lowest pump intensities. (c) The observed fringe contrast as a function of the pump power through the 2nd pass scanned from 200mW (average power) in two ways, either by attenuating the pump before entering the fiber (red squares) or by attenuation after the first pass (blue rhombuses). Below 80mW, a clear difference between the two scenarios is obvious, marking the transition between the quantum regime of single bi-photons at low pump powers and the multi bi-photon semi-classical regime at high powers. (d) Theoretical contrast calculations for both scenarios as a function of average pump power. The slight difference between calculation and experimental results is due to imprecise knowledge of the PCF dispersion.}
\end{figure}

Here, we exploit a nonlinear homodyne interference scheme \cite{Shaked2014Observing} to measure the bi-photon correlation (amplitude and phase) of spontaneously generated FWM and distinguish classical from quantum behavior across the quantum-to-classical transition over a vast dynamic range in intensity ($\sim80dB$). We also demonstrate phenomena of bi-photon generation with imaginary gain, which leads to collapse and revival of the interference contrast. The experimental concept is schematically depicted in figure \ref{fig:setup}(a), where FWM occurs in two separated fibers arranged in series. In between, phase and amplitude manipulations can be performed on the FWM light (including pump), such as dispersion and attenuation. Since all three fields (signal, idler and pump) enter the second fiber, the nonlinear conversion in the second pass can either continue to be from pump to sidebands or be reversed, as dictated by the relative phase between a particular signal-idler pair and the pump ($\phi_s\!+\!\phi_i\!-\!2\phi_p$). If this phase varies spectrally (e.g. due to dispersion of an optical element in the beam), interference fringes appear on the FWM spectrum after the second pass, as shown in figure \ref{fig:contrast784nm}(a). Quantum mechanically, the fringes reflect interference between two indistinguishable possibilities for the creation of a photon - either in the first fiber or in the second, as illustrated in figure \ref{fig:setup}(b).

The bi-photon interference represents a new method to detect quantum correlation that can fully exploit the ultra-high flux of time-energy entangled bi-photons \cite{harris2007chirp,dayan2004two,peer2005temporal,dayan2005nonlinear}, as it measures the spectral correlation at a detection rate comparable to the photon flux \cite{Shaked2014Observing}. One can discern quantum versus classical behavior by comparing the observed fringe-contrast ($ V\!\equiv\!\left(I_{max}\!-\!I_{min}\right)/\left(I_{max}\!+\!I_{min}\right)$) in two experimental scenarios, which are predicted to have a clear discrepancy between classical predictions and those of quantum mechanics. Specifically, one can attenuate the light entering the second fiber in two ways - either by attenuating the pump before the first fiber, which reduces the overall generated FWM sidebands intensity; or by attenuating both the pump and the FWM sidebands between the fibers. Quantum-mechanically, the first scenario attenuates the FWM generation rate, but does not alter signal-idler correlations (two-photon loss), whereas the second scenario is equivalent to the introduction of a beam splitter between the fibers (one-photon loss), which mixes the light with an additional vacuum mode, diminishing the signal-idler correlation. One-photon loss is equivalent in the quantum bi-photon interferometer to an attempt to obtain "which path" information by "stealing" one of the photons, which reduces the interference contrast \cite{englert1966fringe}.

As the gain in the fibers increases, stimulated emission of multiple photon pairs becomes pronounced, and the quantum prediction converges towards the classical one, where both scenarios produce equal interference contrast (see complete quantum and classical analysis in the online supplementary material). We expect therefore that at high-power the contrast will be similar irrespective of how the light was attenuated, but at low power the method of attenuation will reveal the difference between the quantum and classical predictions \cite{shapiro1994semiclassical,herzog1994frustrated,wiseman2000induced,atature2001entanglement,resch2002conditional}, providing a \emph{complete measurement of the classical-to-quantum transition}. The bi-photon interference measurement also exposes a new regime of bi-photon generation, where the parametric gain becomes imaginary. In this regime, FWM generation no longer grows exponentially with fiber length, but is oscillatory, which leads to periodic collapse and revival of the interference contrast.

Below we describe the experimental results, along with quantum and classical calculations. The classical model highlights the failure point of the classical theory, whereas the quantum calculation reconstructs the measured interference visibility across the quantum-to-classical transition. In our experiment, a photonic crystal fiber (PCF) is pumped by narrowband, 6 ps pulses near the zero-dispersion wavelength of the fiber, generating ultra-broadband signal-idler photon pairs ($\sim\!100 $nm signal bandwidth and corresponding idler). The use of picosecond pump pulses generates FWM in a unique regime, where stimulated time-dependent effects, such as self- and cross-phase modulation (SPM and XPM), are weak and limited to the spectral vicinity of the intense pump \cite{Vered2012Two}. The broadband signal-idler fields far from the pump frequency are generated purely by spontaneous FWM, similar to amplified spontaneous emission (ASE) in a laser amplifier. The peak power of the pump pulses provides a large parametric gain, capable of generating in a single pass through the unseeded fiber intense FWM sidebands with many mW of average power \cite{Vered2012Two}.

We measured the fringe visibility as a function of pump power in the two attenuation scenarios as a function of average pump power between 200 mW and 3 mW (corresponding to $~80 dB$ in generated FWM sidebands intensity), as shown in figure \ref{fig:contrast784nm}. At high pump power no significant difference in the fringe contrast is observed between the two attenuation scenarios (classical regime), but below a certain pump power ($\sim 80 mW$), corresponding to the generation of a single photon-pair (on average) per correlation time ($\sim25$ fs), a clear difference is observed, which is most pronounced at the lowest powers. The experimental results are in qualitative agreement with a fully quantum theoretical model (detailed below).

\begin{figure}
\includegraphics[width = 8.6cm]{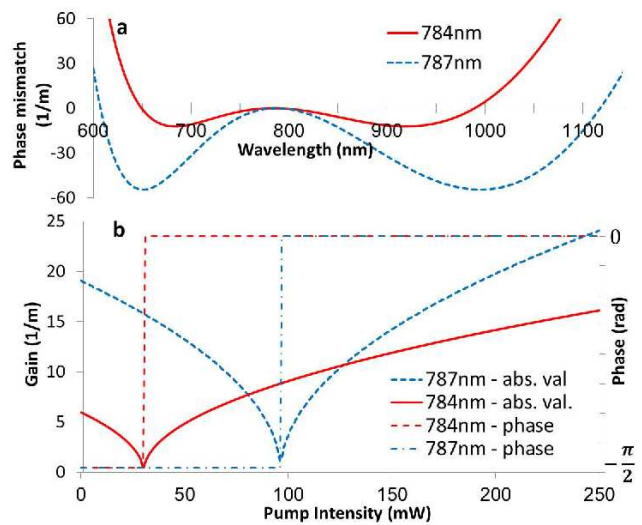}
\caption{\label{fig:mismatch} (a) The calculated bare phase mismatch of our PCF as a function of signal/idler frequency for two pump wavelengths - one very close to the zero dispersion point at 784nm (solid red), and one slightly away from it at 787 nm (dashed blue). (b) Corresponding dependence of the gain amplitude and phase on pump intensity for the same two pump wavelengths, sampled at the center of the signal/idler spectrum, showing the threshold intensity for imaginary gain.}
\end{figure}

It is important now to consider the gain properties of FWM in more detail. The phase mismatch $\Delta q$ depends not only on the dispersion properties of the medium, but also on the pump intensity via XPM, such that $\Delta q\!=\!\Delta k\!+\!2\gamma I_p$, where $\Delta k$ is the 'bare' phase mismatch of the fiber due to dispersion, $I_p$ is the pump intensity and $\gamma$ is the nonlinear Kerr coefficient \cite{Agrawal2000NL, Vered2012Two}. Consequently, the FWM gain is
\begin{equation}
\label{FWMgain}
g\!=\!\sqrt{\gamma^2 I_p^2\!-\!\frac{1}{4}\Delta q^2}\!=\!\sqrt{-\Delta k \gamma I_p\!-\!\frac{1}{4}\Delta k^2}.
\end{equation}
Note that real gain exists only for negative bare phase mismatch $\Delta k\!<\!0$, and that the gain always becomes imaginary when the pump intensity is below a certain threshold $I_p\!<\!\left|\Delta k / 4 \gamma\right|$. FWM with imaginary gain is similar to nonlinear conversion with a phase mismatch: the intensity of the FWM sidebands oscillates with propagation distance $z$ as $I_{FWM}\left(z\right)\!\propto\!\left(I_p/\left| g \right|\right)^2 \sin^2\left( \left| g \right| z \right)$, rather than grow exponentially as $I_{FWM}\left(z\right)\!\propto\!\left(I_p/g\right)^2 \sinh^2\left( gz \right)$. The threshold for imaginary gain is determined by the dispersion properties of the fiber and the pump wavelength, as shown in figure \ref{fig:mismatch}.

\begin{figure}
\includegraphics[width = 8.6cm]{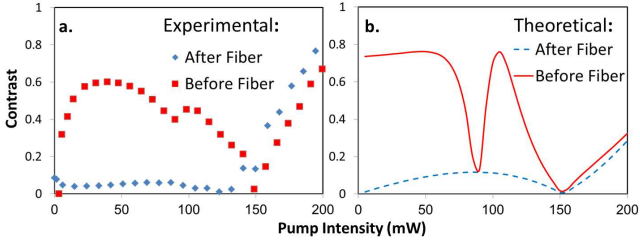}
\caption{\label{fig:contrast787nm} Observation of imaginary gain: (a) The measured fringe contrast as a function of the pump power after the 2nd pass through the fiber, scanned from 250mw down to 5mW (average power) when the pump is tuned slightly away from zero dispersion (784nm). Red squares - pump attenuation before both passes through the fiber (two-photon loss). Blue rhombuses - attenuation of both pump and FWM sidebands between the passes (one-photon loss). At 150mW, the fringe-contrast vanishes for both scenarios, but revives only for the first scenario (two-photon loss). (b) Calculated behavior (see main text for details).}
\end{figure}

While for classical applications the regime of imaginary gain is of little interest due to the low intensity of the generated sidebands, the quantum mechanical properties of the bi-photons are interesting, and have not been previously explored. We thus repeated our measurements at a different pump wavelength, which determines the threshold for imaginary gain, as shown in figure \ref{fig:mismatch}. The results of figure \ref{fig:contrast784nm} were obtained when the pump wavelength was tuned very close to the zero dispersion of the PCF (near 784 nm), where the bare phase mismatch is $\Delta k\approx 0$, and the FWM gain is real at almost any pump intensity. Tuning the pump wavelength slightly further to 787 nm, where the bare phase mismatch is slightly larger, a threshold in pump intensity is expected for real gain values; and indeed an oscillatory collapse-and-revival of the fringe-contrast was observed (figure \ref{fig:contrast787nm}). These oscillations were evident only when the pump intensity was attenuated in front of the first fiber (two-photon loss), whereas for attenuation between the passes (one-photon loss) the contrast collapsed but did not revive. A movie clip is provided in the supplementary material, where the measured spectral fringes are shown as the pump intensity is varied (two-photon attenuation), demonstrating the collapse and revival of the contrast for different spectral regions at different pump intensities. The movie also reflects the intensity dependence of the spectral phase due to XPM, which shifts the entire fringe pattern.

\begin{figure}
\includegraphics[width=8.6cm]{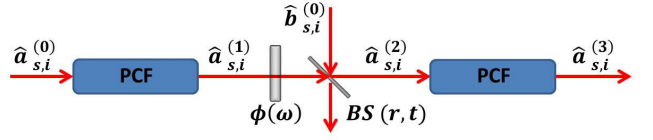}
\caption{\label{fig:calculation} Notation of field operators at the different locations within the experimental setup. $\hat{a}_{s,i}^{(0)}$ are the field operators at the input; $\hat{a}_{s,i}^{(1)}$ are the operators at the output of the first fiber; $\hat{b}_{s,i}^{(0)}$ are the operators of the second input to the beam-splitter (BS); $\hat{a}_{s,i}^{(2)}$ are the operators at the input of the second fiber; and $\hat{a}_{s,i}^{(3)}$ are the operators at the final output.}
\end{figure}

To account for the observed contrast in both the real and imaginary gain regimes, a fully quantum model of the interference was derived under the two loss scenarios (detailed in the supplementary material). We calculated the photon-flux at the output of the second FWM fiber by propagating the field operators (creation and annihilation) through the experimental setup in a three-step propagation (see figure \ref{fig:calculation}): first, through the first FWM fiber, then through the intermediate loss beam-splitter, and last through the second FWM fiber (with appropriate pump phase). The quantum calculation yields the following interference visibility:
\begin{equation}
\label{full_quantum_contrast}
V_{q}\!=\!\frac{2\left | t \right |^2\left |\alpha_1  \right |\left |\beta_1  \right |\left |\alpha_2  \right |\left |\beta_2  \right |}{\left | t \right |^2 \left ( \left |\alpha_2  \right |^2\left |\beta_1  \right |^2\!+\!\left |\alpha_1  \right |^2\left |\beta_2  \right |^2    \right )\!+\!    \left | r \right |^2 \left | \beta_2 \right |^2}
\end{equation}
where $\alpha_j\!=\!\cosh \left ( g_j z \right )\!+\!i \left(\Delta q_j/2g_j\right) \sinh \left( g_j z \right) $, $\beta_j\!=\!i\left(\gamma I_{p_j}/g_j\right) \sinh \left( g_j z \right)$ and $j\!=\!1,2$ refers to the first and second fibers, as the pump intensity and the gain may differ between the fibers, due to the intermediate loss. $t,r$ are the amplitude transmission and reflection coefficients of the assumed intermediate beam-splitter.

Let us examine eq. \ref{full_quantum_contrast} under the simplifying ideal conditions, where the nonlinear interaction is perfectly phase matched ($\Delta q \approx 0$) and the intermediate loss can be neglected ($\left | t \right |^2\approx 1, \left | r \right |^2\approx 0 $). The quantum interference contrast then reduces to
\begin{equation}
\label{quantum_contrast}
V_q=\frac{\sinh\left (2 g_1 z \right )\sinh\left (2 g_2 z \right )}{\cosh\left (2 g_1 z \right )\cosh\left (2 g_2 z \right )\!-\!1}
\end{equation}
This result should be compared to the classical prediction (derived in the supplementary material)
\begin{equation}
\label{classical_contrast}
V_c\!=\!\frac{\sinh\left (2 g_1 z \right )\sinh\left (2 g_2 z \right )}{\cosh\left (2 g_1 z \right )\cosh\left (2 g_2 z \right )}\!=\!\tanh\left (2 g_1 z \right )\tanh\left (2 g_2 z \right ),
\end{equation}
which indicates that the classical model is valid at high FWM gain ($\cosh\left ( g_1 z \right )\cosh\left ( g_2 z \right )\!\gg\!1$), when stimulation dominates, but breaks down when spontaneous FWM is substantial (the average number of generated bi-photons per correlation time is of order unity).

The theoretical predictions in figures \ref{fig:contrast784nm} (d) and \ref{fig:contrast787nm} (b) were all calculated with equation \ref{full_quantum_contrast} while varying either the intermediate attenuation $t$ or the pump intensity in front of the first fiber. Note that in the actual experiment some intermediate loss was always present even with no intermediate beam splitter, due to recoupling losses  back into the fiber. Consequently, the two passes are not exactly symmetric (the pump intensity in the second pass through the fiber is approximately $75\%$ of the intensity in the first pass), and FWM generation with imaginary gain vanishes in each fiber at a different pump intensity. The collapses of the fringe-contrast match the situations where generation is nulled in one of the fibers, but not the other. The calculation qualitatively reconstructs the features of collapse and revival in the fringe-contrast, though differences in the exact shape and absolute values of the contrast are evident. The reasons for this discrepancy are assumed to be: first, the assumption of a CW single-frequency pump in the calculation neglects temporal effects associated with the time variation of the pump intensity (pulse shape) that 'smear out' some of the predicted high-contrast features; second, the calculation is sensitive to the dispersion properties of the fiber that were not precisely known; and last, imprecision of our contrast estimation procedure, which extracted the fringe contrast from the spectral pattern directly. Thus, a dark fringe is measured at a slightly different wavelength than a bright fringe, which may introduce errors if the contrast at different wavelengths is slightly different. This may account for the reduction of the observed contrast at the very low pump intensities, which is not predicted by the theory.

In conclusion, we presented a simple two-photon interference method for investigating the quantum correlation of broadband bi-photons generated by FWM which can be applied at any photon flux, from truly single photons to classical power levels. The high gain in our PCF fiber, enabled the observation of the full transition between quantum and classical regimes, and bi-photon generation with imaginary gain was demonstrated and explored.

This research was supported in part by  the EU-IRG program (grant no. 248630).

\end{document}